\documentclass[aps,prd,twocolumn,superscriptaddress,preprintnumbers,floatfix,nofootinbib,longbibliography,showlabels]{revtex4}

\usepackage[utf8x]{inputenc}
\usepackage{graphicx}
\usepackage[dvips]{color}
\usepackage{hyperref}
\usepackage{epsfig}
\usepackage{amsmath}
\usepackage{amssymb}
\newcommand{\tr}{\textrm{Tr}\,}

\newcommand{\msbar}{\overline{\mbox{\rm MS}}}

\def\lsim{\raise0.3ex\hbox{$<$\kern-0.75em\raise-1.1ex\hbox{$\sim$}}}
\def\gsim{\raise0.3ex\hbox{$>$\kern-0.75em\raise-1.1ex\hbox{$\sim$}}}

\newcommand{\beq}{\begin{eqnarray}}
\newcommand{\eeq}{\end{eqnarray}}
\newcommand{\be}{\begin{equation}}
\newcommand{\ee}{\end{equation}}

\newcommand{\bmat}{\left (\begin{array}{cc}}
\newcommand{\emat}{\end{array} \right )}

\newcommand{\susk}{\chi_{\phi^\dagger\phi}}
\newcommand{\suskfit}{\chi_{\rm fit}}

\definecolor{grey}{RGB}{100,100,100}

\newcommand{\half}{\mbox{${\frac12}$}}

\newcommand{\eqr}[1]{(\ref{#1})}
\usepackage{ulem}

\begin{document}

\title{The Standard Model cross-over on the lattice}

\author{Michela D'Onofrio}
\affiliation{Institute for Theoretical Physics,
  Albert Einstein Center, University of Bern,
  Sidlerstrasse 5, CH-3012 Bern, Switzerland}
\affiliation{Department of Physics and Helsinki Institute of Physics, PL 64 (Gustaf H\"allstr\"omin katu 2), FI-00014 University of Helsinki, Finland}

\author{Kari Rummukainen}
\affiliation{Department of Physics and Helsinki Institute of Physics, PL 64 (Gustaf H\"allstr\"omin katu 2), FI-00014 University of Helsinki, Finland}

\date   {\today}

\begin{abstract}
  With the physical Higgs mass the Standard Model symmetry restoration phase transition is a smooth cross-over.  We study the thermodynamics of the cross-over using numerical lattice Monte Carlo simulations of an effective SU(2)$\times$U(1) gauge + Higgs theory, significantly improving on previously published results.  We measure the Higgs field expectation value, thermodynamic quantities like pressure, energy density, speed of sound and heat capacity, and screening masses associated with the Higgs and $Z$ fields.
While the cross-over is smooth, it is very well defined with a width of only $\sim 5$\,GeV.  We measure the cross-over temperature from the maximum of the susceptibility of the Higgs condensate, with the result $T_c = 159.5 \pm 1.5$\,GeV.  Outside of the narrow cross-over region the perturbative results agree well with non-perturbative ones. 
\end {abstract}

\maketitle

\section{Introduction} 

The LHC particle accelerator has spectacularly confirmed the Standard Model of the particle physics: the current combined results from ATLAS and CMS experiments point towards a Higgs boson with a mass $125.1 \pm 0.3$\,GeV \cite{Aad:2015zhl}, and no direct experimental evidence of beyond the Standard Model physics has been observed.

If the electroweak scale physics is fully included in the Standard Model, we are in the position to have a complete description of the high-temperature electroweak phase transition between the low-temperature broken phase, where the Higgs field has a non-vanishing expectation value, and the high-temperature symmetric phase.\footnote{%
Although there is no real symmetry breaking phase transition, we use the conventional labels ``broken'' and ``symmetric'' to refer to the low- and high-temperature phases, respectively.}   
The overall nature of the transition was settled already in 1995--98 using lattice simulations \cite{Kajantie:1996mn,Karsch:1996yh,Gurtler:1997hr,Rummukainen:1998as,Csikor:1998eu,Aoki:1999fi}, which indicate a first-order phase transition for Higgs masses $\lsim\,72$\,GeV, and a smooth cross-over otherwise.  Most of these simulations were done using an effective 3-dimensional theory of the full Standard Model, derived using perturbation theory.
The physics of the transition is non-perturbative due to the infrared singularities arising at momentum scales $\lsim g^2 T$ \cite{Linde:1978px,Gross:1980br}.  These modes are fully captured in effective 3-dimensional theories \cite{Ginsparg:1980ef,Appelquist:1981vg,Nadkarni:1982kb,Landsman:1989be,Jakovac:1994xg,Farakos:1994xh,generic,Braaten:1995cm}, which provide an economical and accurate way to study the non-perturbative physics at the cross-over.

A cross-over means that the early Universe evolved smoothly from symmetric to broken phase without deviating significantly from the thermal equilibrium, rendering e.g.\, electroweak baryogenesis \cite{Kuzmin:1985mm,Rubakov:1996vz} ineffective.  Nevertheless, the cross-over could still influence other processes going on at the same time.  An important example is the rate of B+L violation, sphaleron rate, which switches off somewhat below the cross-over temperature \cite{D'Onofrio:2014kta,D'Onofrio:2012jk} and may affect the baryon number generation in leptogenesis \cite{Asaka:2005pn,Canetti:2012kh}.  Decoupling of the dark matter may also be affected by the details of the Standard Model equation of state in certain scenarios \cite{Steigman:2012nb,Hindmarsh:2005ix}.

The Standard Model cross-over with the physical Higgs mass of $125$\,GeV was recenty studied on the lattice in ref.~\cite{D'Onofrio:2014kta}.  The cross-over temperature was reported to be $T_c = 159.5 \pm 1.5$\,GeV.  However, the focus of this study was the sphaleron rate as a function of the temperature, and for quantities related to thermodynamics the results were limited: as an example, the Higgs field expectation value $\langle \phi^\dagger \phi \rangle $ was determined only at one lattice spacing, and the effects of the hypercharge U(1) field were neglected.  

Recently, Laine and Meyer \cite{Laine:2015kra} used up to 3-loop perturbative computations and existing results \cite{Gynther:2005dj,Gynther:2005av} to derive a relation between the trace anomaly of the Standard Model and the Higgs field expectation value in the 3-dimensional effective theory.  Using the numerical results from ref.\,\cite{D'Onofrio:2014kta}, they calculated several thermodynamic quantities across the cross-over.  The combination of perturbative calculations and lattice simulations avoids the infrared problems which make purely perturbative analysis unreliable near the cross-over.
The set-up is related to the computation of the QCD pressure using effective theory \cite{Hietanen:2008tv}; however, for the Standard Model electroweak sector the couplings are smaller and the method can be expected to have much better accuracy.

In this work our goal is to improve on the state of the art in lattice simulations of the Standard Model cross-over: we include the U(1) gauge field, and we use several lattice spacings and large volumes, enabling us to do reliable continuum extrapolation.  The model we simulate is an effective three-dimensional theory containing SU(2)$\times$U(1) gauge fields and a Higgs scalar.  This particular model has been used previously to study the Standard Model phase transition \cite{su2xu1}, also with an external hypermagnetic field \cite{Kajantie:1998rz}.  However, these older studies used unphysical Higgs masses.

As discussed above, measuring the Higgs condensate is essential for obtaining non-perturbative contributions to thermodynamic quantities.  Thus, we pay special attention to precise measurement of the gauge invariant expectation value $\langle \phi^\dagger \phi\rangle$ and its susceptibility.  The maximum of the susceptibility allows us to determine the pseudocritical temperature accurately, at $T_c = 159.6 \pm 0.1 \pm 1.5$\,GeV.  The first error
is due to the precision of the lattice simulations and the second is a conservative estimate of the uncertainty of the effective theory description \cite{generic}.   The pseudocritical temperature is completely consistent with the result in ref.\,\cite{D'Onofrio:2014kta}.

We use the condensate and the susceptibility to obtain several thermodynamic quantities across the cross-over: energy density, pressure, heat capacity, speed of sound, and equation of state parameter.  For most quantities the magnitude of the effects of the cross-over are small, only at a percent level, but nevertheless clearly visible.  We also determine the Higgs and $W^3$ screening masses and the $\gamma-Z$ mixing.  The emerging picture is fully consistent with a smooth and regular cross-over.  The cross-over region, where observables deviate significantly from low- or high-temperature behaviour, is remarkably narrow, between 157 and 162 GeV.

This paper is organized as follows.  In section \ref{sec:efftheory} we describe the effective theory and in section \ref{sec:lattice} its implementation on the lattice.  The Higgs condensate and its susceptibility are discussed in section \ref{sec:condensate}, fundamental thermodynamic observables in section \ref{sec:thermo} and the screening masses in section \ref{sec:screening}.  We conclude in section \ref{sec:conclusions}.

\section{Effective three-dimensional description}
\label{sec:efftheory}


At temperatures of order $100$\,GeV the gauge couplings in the Standard Model are small,
and the Euclidean path integral contains a parametric hierarchy of energy scales: $\pi T$, $gT$ and $g^2T$.  While the harder scales can be reliably treated with perturbation theory, at $k\sim g^2 T$ we have to face the non-perturbative infrared physics \cite{Linde:1978px}.  The non-perturbative physics can be captured into an effective theory for the soft $k\sim g^2 T$ scales, obtained by integrating over the harder scales using well-defined perturbative methods.  This effective
theory is purely bosonic and three-dimensional.
%
The detailed description of the derivation of the theory can be found in refs.\,\cite{generic,Farakos:1994xh}.

\begin{figure}[t]
  \includegraphics[width=0.95\columnwidth]{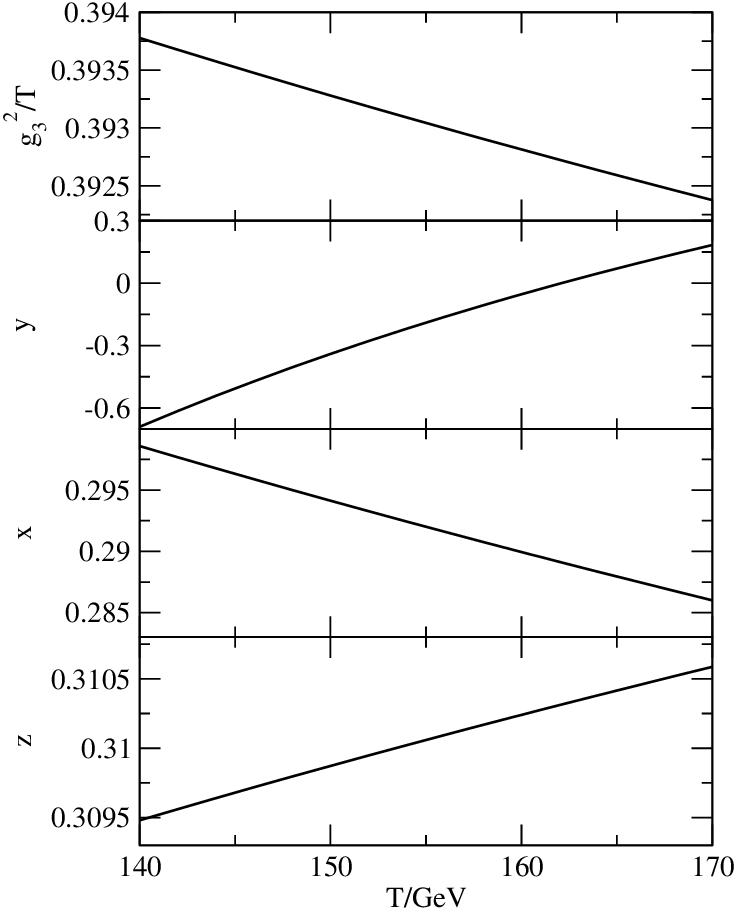}
  \caption{The parameters of the effective theory \eqr{contaction} as
    functions of the physical temperature.}
  \label{fig:params} 
  \vspace{-0mm}
\end{figure}

The Lagrangian of the effective theory is the 3-dimensional SU(2)$\times$U(1) gauge theory with a Higgs field%
\begin{align}
L =& \frac 1 4 F_{ij}^aF_{ij}^a+ \frac 1 4 B_{ij}B_{ij} \nonumber \\
&+ (D_i\phi)^\dagger D_i\phi+m_3^2\phi^\dagger\phi + \lambda_3(\phi^\dagger\phi)^2,
\label{contaction}
\end{align}
where
\begin{align}
F_{ij} &= \partial_iA_j -\partial_jA_i - g_3
[A_i, A_j],  ~~~
A_i =  \half\sigma_aA_i^a \nonumber\\
B_{ij} &= \partial_iB_j-\partial_jB_i    \\
D_i&= \partial_i+ig_3A_i+ig'_3B_i/2.  \nonumber
\end{align}
Here $A_i$ and $B_i$ are the 3-dimensional SU(2) and U(1) gauge fields, $g_3^2$ and ${g'_3}^2$ the dimensionful SU(2) and U(1) couplings, and $\phi$ a complex doublet.
The SU(2)$\times$U(1) local gauge transformation is
\begin{equation}
\phi(x)\to e^{i\alpha(x)}G(x)\phi(x).
\label{phitransf}
\end{equation}
The parameters appearing in the effective theory are
dimensionful.  If we take one of the parameters, say $g_3^2$, to set the
scale, the dynamics
then depends on the three dimensionless parameters $x$, $y$ and $z$,
defined as
\begin{equation}
x\equiv \frac{\lambda_3}{g_3^2},\qquad
y\equiv \frac{m_3^2}{g_3^4},\qquad
z\equiv \frac{g_3'^2}{g_3^2}.
\label{3dvariables}
\end{equation}
The four parameters  $g_3^2$, $g_3'^2$, $\lambda_3$
and $m_3^2$ are definite perturbatively computable functions of the Standard Model
parameters ($\alpha_S(M_W)$, $G_F$, $M_{\rm Higgs}$, $M_W$, $M_Z$, $M_{\rm top}$), and the temperature $T$. These have been computed through a set of 1- and 2-loop matching relations \cite{generic}, and are shown in figure \ref{fig:params} as functions of the temperature. 
The accuracy of the effective theory can be estimated to be at $\sim 1\%$  level, as discussed in refs.\,\cite{generic,Laine:1999rv}.

From figure \ref{fig:params} we see that only $y$ has large temperature dependence.  Indeed, from the effective theory point of view it is natural to choose $y$ as the temperature variable, although we present our results in terms of the physical temperature.  The transition is expected to happen near $y=0$, which occurs at $T=162.1$\,GeV.

\section{Lattice action and simulations}
\label{sec:lattice}

For the lattice implementation it is convenient to introduce a matrix
parametrisation of the Higgs field by writing
\begin{equation}
\Phi= \frac{1}{g_3^2} \biggl((\tilde\phi)(\phi)\biggr)\equiv
\frac{1}{g_3^2} 
\left(\begin{array}{cc}
    \phi_2^*&\phi_1\\
    -\phi_1^*&\phi_2
  \end{array}\right).
\label{phimatrix2}
\end{equation}
Under an SU(2)$\times$U(1) gauge transformation $\Phi$
transforms according to
\begin{equation}
\Phi(x)\to G(x)\Phi e^{-i\theta(x)\sigma_3}.
\label{gt}
\end{equation}
The lattice action corresponding to the continuum theory
\eqr{contaction} is
\begin{align}
S&= \beta_G \sum_x \sum_{i<j}[1-\half \tr P_{ij}] 
  +  \frac{\beta_G}{z} \sum_x \sum_{i<j} \half \alpha_{ij}^2 \nonumber\\
 &- \beta_H \sum_x \sum_{i}
\half\tr\Phi^\dagger(x)U_i(x)\Phi(x+i)
e^{-i\alpha_i(x)\sigma_3}   \label{lagrangian}  \\
 &+ \beta_2 \sum_x
 \half \tr\Phi^\dagger(x)\Phi(x) + \beta_4\sum_x
 \big[\half\tr\Phi^\dagger(x)\Phi(x)\big]^2. \nonumber
\end{align}
Here the SU(2) and the (non-compact) U(1) plaquettes are
\begin{align}
  P_{ij}(x) &= U_i(x)U_j(x+\hat i)
  U^\dagger_i(x+\hat j)U^\dagger_j(x), \label{su2plaq} \\
  \alpha_{ij}(x) &=\alpha_i(x)+\alpha_j(x+\hat i)-
  \alpha_i(x+\hat j)-\alpha_j(x). \label{u1plaq}
\end{align}
The parameters of the lattice action
$\beta_G$, $\beta_H$, $\beta_2$ and $\beta_4$ can be expressed
in terms of the continuum parameters and lattice spacing $a$ 
using relations \cite{Kajantie:1998rz,Laine:1997dy}
\begin{align}
  \beta_G  &= \frac 4 {g_3^2a}, \label{latcont1} \\
  \beta_H  &= \frac 8 {\beta_G}, ~~~
  \beta_4  =  \frac{\beta_H^2}{\beta_G}, \\
  \frac{\beta_2}{\beta_H}  &=   3 + \frac{8y}{\beta_G^2}
            - (3 + 12 x + z)\frac{\Sigma}{4\pi\beta_G} \nonumber \\
            &-\frac{1}{2\pi^2\beta_G^2} \bigg[ 
             \left(
               \frac{51}{16} - \frac{9z}8 - \frac{5z^2}{16} 
               +9x - 12x^2 + 3xz \right)  \nonumber \\
            & ~~~~~~~\times\big(\log(\frac{3\beta_G}2) + 0.09\big) \nonumber \\
           &+ 5 - 0.9z + 0.01z^2 + 5.2x + 1.7xz\bigg],
            \label{latcont2}
\end{align}
where $\Sigma=3.17591$. These relations become exact in the limit
$a\rightarrow 0$ ($\beta_G \rightarrow \infty$), with an error proportional to $a$.  In refs.\,\cite{Moore:1996bf,Moore:1997np} partial $O(a)$ improvement of the relations was presented; however, for simplicity we do not implement
this here.

Finally,
the gauge-invariant lattice observable
$\langle\frac12\tr\Phi^\dagger\Phi\rangle$
is related to the renormalized 3-dimensional continuum
quantity $\langle\phi^\dagger\phi\rangle$ in the $\msbar$ scheme 
by
\begin{equation}
{\langle\phi^\dagger\phi\rangle\over g_3^2}=
\langle \half\tr\Phi^\dagger\Phi \rangle-
\frac{\Sigma\beta_G}{8\pi}-
\frac{3+z}{16\pi^2}\biggl(\log( 3\beta_G/2)
+0.6678\biggr).\label{rl2}
\end{equation}
This relation again has corrections at order $O(a)$.
The 3-dimensional condensate is related
to the physical Standard Model 
Higgs condensate $v$ by
\begin{equation}
  v^2/T^2 = 2\langle \phi^\dagger \phi \rangle/T .
\end{equation}

We use 3 lattice spacings, $a g_3^2 = 4/\beta_G$, with $\beta_G = 6$, $9$ and $16$, with volumes up to $(120 a)^3$, as listed in table~\ref{tab:sims}.  
The measurements are done at 24 temperature values between 140 and 170\,GeV.
The simulation algorithm is a mixture of overrelaxation and heat bath update steps, described in detail (without a U(1) field) in \cite{nonpert}.  The measurements of the observables are done after two repetitions of
4 full overrelaxation update sweeps followed by one heat bath update.  The number of measurements varies between 10\,000 and 350\,000, with most of the measurements done near the cross-over temperature.

\begin{table}[ht]
\begin{tabular}{|c|l|}
\hline
$\beta_G = 4/g_3 a$ & volumes/$a^3$ \\
\hline
 6                  & $24^3$, $48^3$ \\
 9                  & $24^3$, $32^3$, $60^3$ \\
16                  & $56^3$, $120^3$ \\
\hline
\end{tabular}
\caption{The lattice spacings and sizes used in the analysis.  For each lattice,
  we use 24 temperature values between 140\,GeV and 170\,GeV\@.  The number of measurements varies between 10\,000 -- 400\,000, with the largest number of measurements near the cross-over temperature.}
\label{tab:sims}
\end{table}

\section{Higgs condensate and the pseudocritical temperature}
\label{sec:condensate}

\begin{figure}
  \includegraphics[scale=0.6]{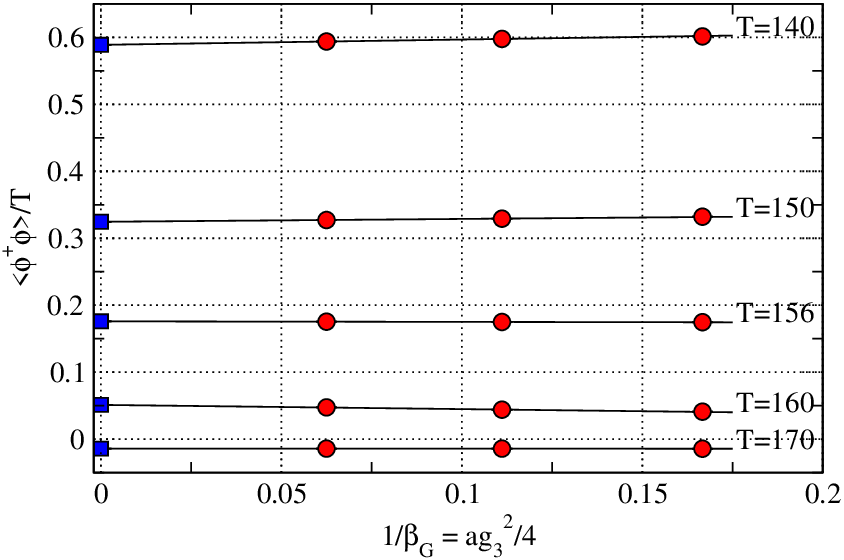}
  \caption{The continuum limit of $\langle \phi^\dagger\phi\rangle$
    at a few selected temperature values.  The statistical errors are too small to be visible at this scale.}
  \label{fig:cont_phi2}
\end{figure}

The measurements of the Higgs condensate $\langle \phi^\dagger \phi \rangle$ are extrapolated to the continuum with linear in $a$ extrapolation,
as shown in figure \ref{fig:cont_phi2} at a few selected temperatures.
For this quantity the finite volume effects are negligible, indistinguishable from the statistical accuracy of the measurements, and we use the largest volumes in our analysis.  Below we analyze the finite volume effects of the susceptibility of $\phi^\dagger\phi$ in more detail and show that these also remain very small.
Given the presence of the massless U(1) field, the smallness of the finite volume effects is not a priori obvious, but is in agreement with the earlier results in \cite{su2xu1}.

\begin{figure}
  \includegraphics[scale=0.6]{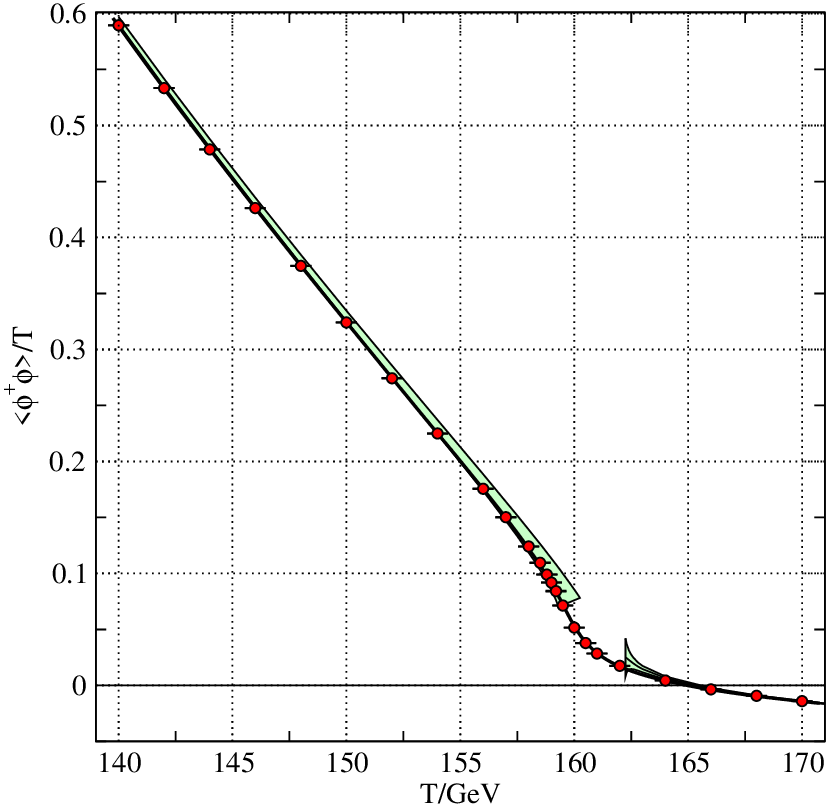}
  \caption{The continuum result of $\langle\phi^\dagger\phi\rangle$,
    compared with the perturbative broken and symmetric phase results.  The shaded bands are estimations of unknown higher order corrections to perturbative results.  The solid continuous line is an interpolation to the data.}
  \label{fig:phi2}
\end{figure}

The continuum extrapolation of $\langle\phi^\dagger\phi\rangle$ is shown in figure \ref{fig:phi2}.
We compare the numerical result with the perturbative broken phase 2-loop Coleman-Weinberg computation \cite{nonpert} and the symmetric phase 3-loop result \cite{Laine:2015kra}.  In both cases the higher order contributions are estimated with shaded bands.  The agreement between the perturbative results and the lattice results is remarkably good, especially in the symmetric phase where the perturbative expansion converges quickly.\footnote{%
  Figure \ref{fig:phi2} can be compared with figure 2 in ref.\,\cite{Laine:2015kra}, where the agreement between the lattice and the perturbative results is much weaker,  due to the missing continuum limit of the lattice results.}
There is only a narrow window of a few GeV around the cross-over temperature (corresponding to $y\approx 0$) where the perturbative expansions do not converge.  

The apparent good convergence in the symmetric phase may be surprising, because in this phase the non-abelian gauge bosons are perturbatively massless, making the physics at soft momentum scales $k\sim g²T$ non-perturbative \cite{Linde:1978px}.  The excellent match between the lattice and the perturbation theory means that for the Higgs condensate their effect remains small.  This can be contrasted with e.g.\,the sphaleron rate, which is in essence completely determined by the soft physics.

\begin{figure}
  \includegraphics[scale=0.6]{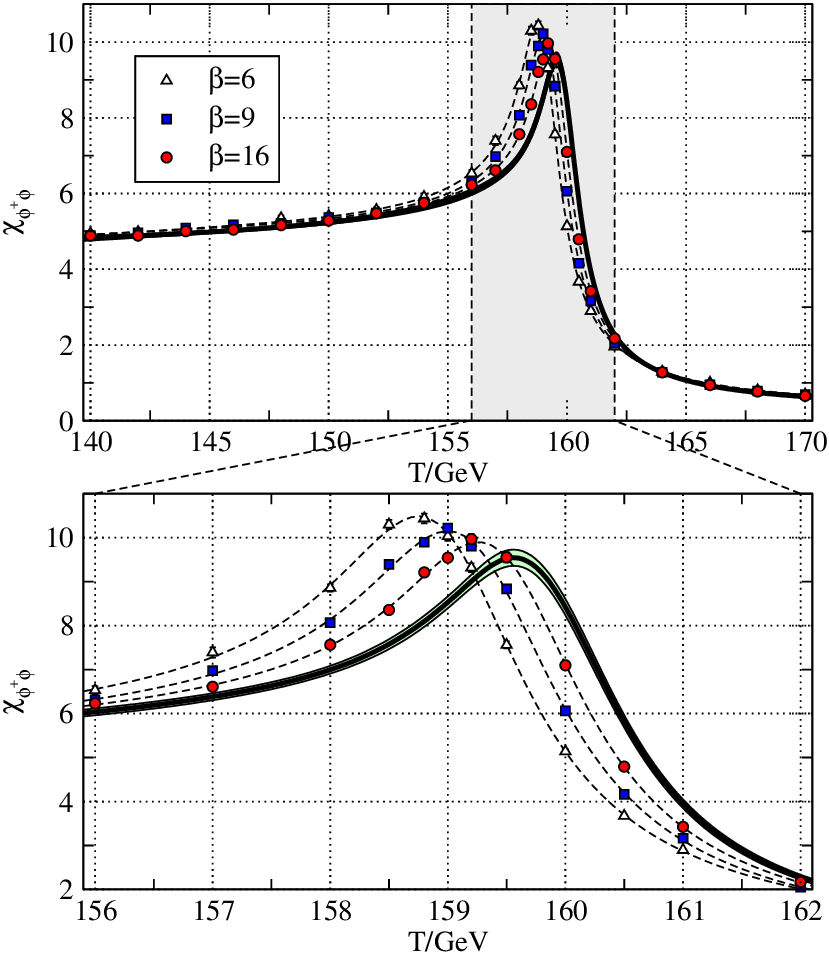}
  \caption{ Above: susceptibility $\susk$
    shown at $\beta_G=6$, $9$ and
    $16$, together with the interpolating functions.
    The continuum limit is shown
    with a heavy line.  Below: As above, zoomed-in to the shaded band near the cross-over region.}
  \label{fig:phi22}
\end{figure}

We define the pseudocritical temperature by the maximum location of the dimensionless susceptibility
\begin{equation}
  \susk
  = V T \left\langle [(\phi^\dagger\phi)_V - \langle (\phi^\dagger\phi)_V\rangle ]^2 \right\rangle,
\end{equation}
where $(\phi^\dagger\phi)_V = 1/V \int dV \phi^\dagger\phi$ is the volume average of $\phi^\dagger\phi$.
This is shown in figure \ref{fig:phi22}, for the largest simulation volumes at each lattice spacing.  The use of the largest volumes is justified below.  There is a well-defined peak near the cross-over temperature,
however, the location of the peak has a clear lattice spacing dependence.  Because of the narrowness of the peak, the continuum limit extrapolation becomes delicate: at a fixed temperature, the values of $\susk$ at different lattice spacings have large and non-uniform variation, which can be clearly seen in the zoomed-in subplot in figure \ref{fig:phi22}.  Now a linear or a linear + quadratic in $a$ continuum extrapolation at fixed temperature does not give a reasonable result using the available lattice spacings.

We obtain a much better controlled continuum limit if we first ``undo'' the shift in the temperature so that the peaks become aligned.  This is perfectly compatible with $O(a)$ lattice spacing effects, if the shift vanishes as $O(a)$ or faster as the continuum limit is approached.  However, shifting the temperature requires that we interpolate the measurements of $\susk$ to continuous functions of $T$.
%
We achieve this by fitting the measurements at each $\beta_G$ to a differentiable interpolating function, $\suskfit(T,\beta_G)$.
The fit procedure and the choice of the function is described in appendix \ref{app:fit}.  The fitted functions at each $\beta_G$ are included in figure \ref{fig:phi22}.  We note that commonly used reweighting methods do not work well in this case, because there is limited overlap in the probability distributions between neighbouring points over the large temperature range.

\begin{figure}
  \includegraphics[scale=0.6]{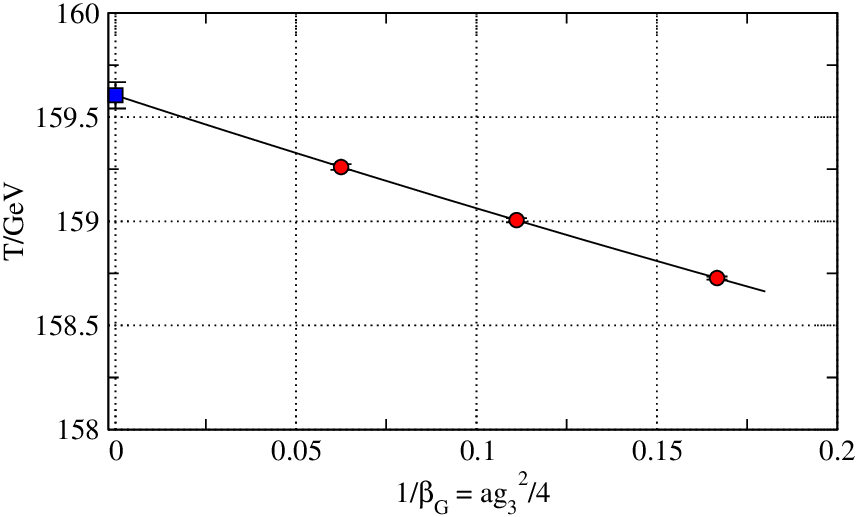}
  \caption{Continuum extrapolation of the maximum location of $\susk$.}
  \label{fig:maxloc_phi22}
\end{figure}

We now define the pseudocritical temperature at each $\beta_G$ by the location of the maxima of the fit functions $\suskfit(T,\beta_G)$.  These are shown in figure \ref{fig:maxloc_phi22} against $1/\beta_G = a g_3^2/4$.  The measurements fall on a remarkably straight line, allowing us to obtain the continuum limit pseudocritical temperature,
\begin{equation}
  T_{\chi_{\rm max}}(\beta_G) = T_c + \frac{C}{\beta_G},
\end{equation}
with the result $T_c = 159.58 \pm 0.06$\,GeV.  The error bars conservatively include the results from both linear in $a$ and linear + quadratic extrapolation.  However, these errors are completely overwhelmed by the per cent-level uncertainties associated with the accuracy of the 3-dimensional effective theory description of the Standard Model.

We use the above fit to shift the temperature variable of the functions $\suskfit$ as $T\rightarrow  T-C/\beta_G$. Now we obtain the continuum limit of $\susk$ by making a linear extrapolation independently at each shifted value of $T$.  The final result of the extrapolation is shown in figure \ref{fig:phi22} as a shaded band.  The errors are propagated using the jackknife method.  As can be observed, this method of extrapolation works very well.  As an aside, if the fully $O(a)$ improved lattice-continuum relations corresponding to equations (\ref{latcont1}--\ref{latcont2}) were available, the $O(a)$ shift in $T$ should be cancelled automatically.

\begin{figure}
  \includegraphics[scale=0.6]{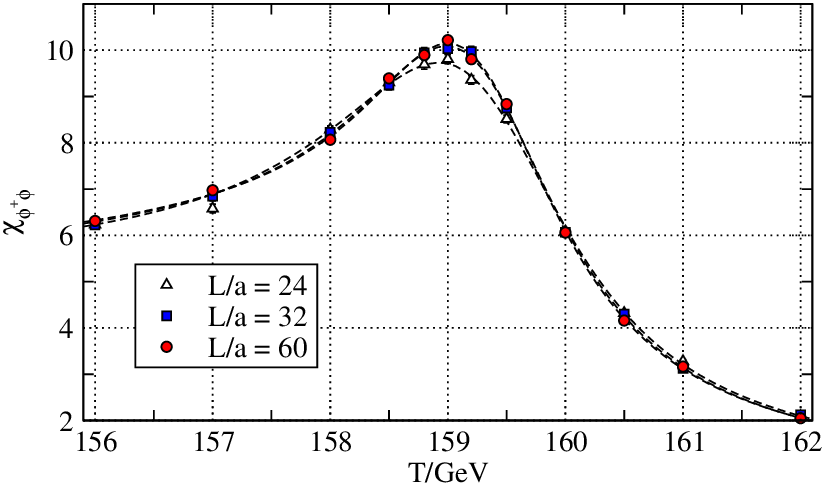}
  \caption{The finite volume behaviour of the susceptibility $\susk$,
    shown for the lattices of size $L/a=24$, $32$ and $60$ at $\beta_G=9$.  Only the smallest volume shows appreciable deviation, which is much below the lattice spacing effects in figure \ref{fig:phi22}.}
  \label{fig:phi22vol}
\end{figure}

Finally, in figure \ref{fig:phi22vol} we show the behaviour of the susceptibility at different volumes at $\beta_G=9$.  In this case we have 3 volumes available, with physical linear lattice sizes $L \approx 27/T$, $36/T$ and $68/T$ (at $L/a=24$, $32$ and $60$, respectively).  The larger volumes agree within statistical errors, and the smallest volume deviates slightly only at the susceptibility peak.  Compatible behaviour is seen at other lattice spacings, where the smallest volumes have $L \gsim 36/T$.  Thus, we conclude that the largest volumes have negligible finite volume corrections for the Higgs condensate and its susceptibility.

\begin{figure*}
\begin{centering}
  \includegraphics[scale=0.6]{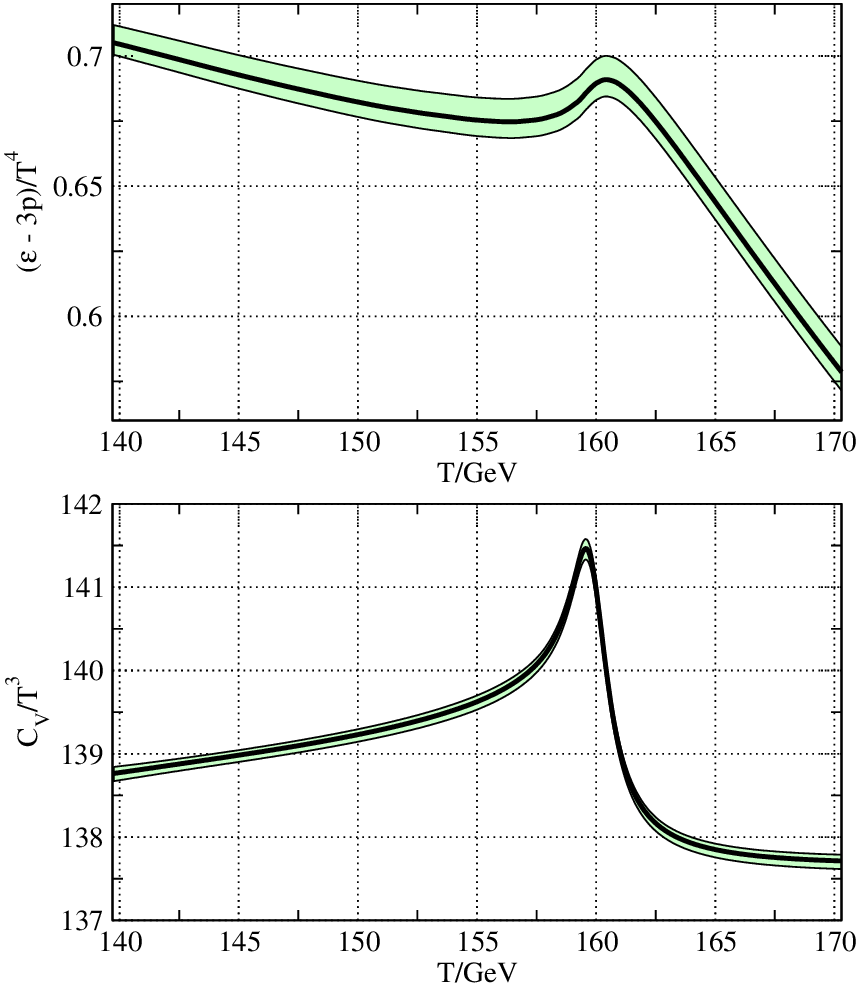}
  \includegraphics[scale=0.6]{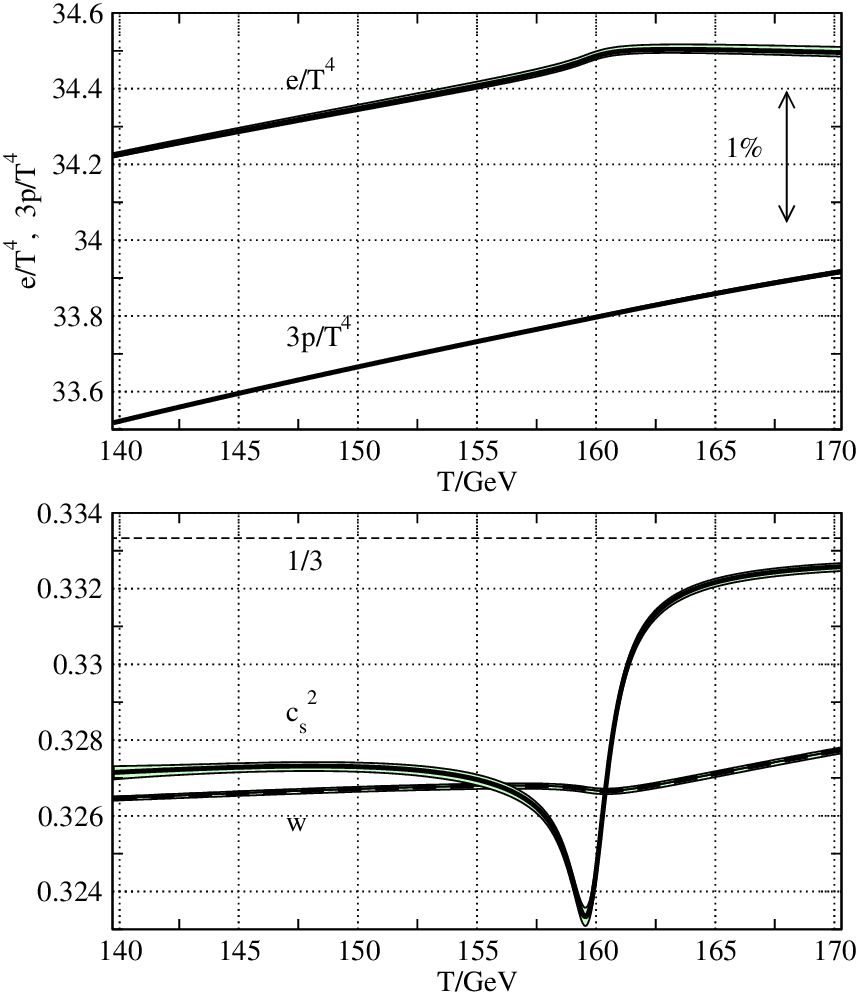}
\end{centering}
\caption[a]{The interaction measure $\Delta=(e-3p)/T^4$ (top left);
  energy density $e$ and pressure $p$ (top right); heat capacity $C_V = d e/dT$ (bottom left);
  speed of sound squared $c_s^2$ and the equation of state parameter $w = p/e$ (bottom right).  The error bands are a combination of the statistical errors and renormalizaton scale variation $\bar\mu=(0.5\ldots 2)\pi T$.
  The energy density and the pressure are affected by a systematic uncertainty of order 1\%, indicated with
  a vertical arrow.}
\label{fig:thermo}
\end{figure*}

In the absence of massless modes the finite volume effects vanish exponentially in $L$.  In this case the system has a massless U(1) gauge boson, and thus one could expect to observe power law finite volume effects.  However, the hypercharge U(1) field participates weakly to the transition, only by mixing with $W^3$ by the Weinberg angle.  In part, this explains the smallness of the finite volume effects.\footnote{%
  Strong external hypermagnetic fields do nevertheless affect the transition, see e.g. \cite{Kajantie:1998rz}.}  Similar behaviour has been observed in earlier lattice studies \cite{su2xu1}.

In order to facilitate the detailed analysis of thermodynamic functions we also interpolate the $\phi^\dagger\phi$-measurements.  We do this by interpolating at each lattice spacing using a spline fit, shifting the temperatures by the same amount as for $\susk$ and extrapolate to continuum.   The result is included in figure \ref{fig:phi2} as a continuous line.  The temperature shift before the continuum extrapolation changes the result by a very small amount which not distinguishable in the figure.  For possible future use the interpolated results for $\phi^\dagger\phi$ and $\susk$, together with other thermodynamic measurements, can be downloaded from
\cite{sm_data}.


\section{Thermodynamics of the cross-over}
\label{sec:thermo}

Recently, Laine and Meyer \cite{Laine:2015kra} showed how one can combine perturbative calculations and effective theory simulation to obtain the Standard Model pressure, energy density and other thermodynamic quantities derived from these.  As an input they used the simulation results from ref.\,\cite{D'Onofrio:2014kta}.  Because these results use only a single lattice spacing and ignore the hypercharge U(1), we revisit the calculation using our improved data.

The fundamental thermodynamic quantitity is the interaction measure (``trace anomaly'')
\begin{equation}
  \Delta \equiv \frac{e(T)-3p(T)}{T^4} = 
  T\frac{\mathrm{d}}{\mathrm{d}T} \frac{p(T)}{T^4}.
\end{equation}
In ref.\,\cite{Laine:2015kra} the interaction measure is split into
three parts: $\Delta = \Delta_1 + \Delta_2 + \Delta_3$, where
$\Delta_1$ includes effects from breaking of scale invariance by
quantum corrections, $\Delta_2$ effects from the Higgs condensate, and
$\Delta_3$ comes from vacuum subtraction.  For our purposes it is convenient to express it as
\begin{equation}
  \Delta(T,\bar\mu) = A(T;\bar\mu) + B(T;\bar\mu) 
  \frac{\langle\phi^\dagger\phi\rangle(T)}{T}
  \label{delta}
\end{equation}
where $\bar\mu$ is the $\overline{\mbox{\small MS}}$ renormalization scale 
and the functions $A$ and $B$ can be computed following ref.\,\cite{Laine:2015kra}, giving $\Delta$ up to parametric order $g^5$.
However, here we use the $O(g^4)$ expression for the function $A$,
because it has been argued that the $O(g^5)$ contribution,
which is mostly due to QCD contributions,
leads to an underestimate of the pressure and the energy density \cite{Laine:2015kra}.  The resulting $\Delta$ is shown in figure \ref{fig:thermo}.

From figure \ref{fig:phi2} we can see that the direct perturbative computation of $\langle\phi^\dagger\phi\rangle(T)$ 
fails to converge near the cross-over temperature. On the other hand, the functions $A(T;\bar\mu)$ and $B(T;\bar\mu)$ do not suffer from this problem.  Therefore, by measuring the Higgs condensate non-perturbatively on the lattice, we obtain well-defined expressions for thermodynamic quantities across the cross-over.

The pressure is obtained from $\Delta$ by integration:
\begin{equation}
  \frac{p(T)}{T^4} -   \frac{p(T_0)}{T_0^4} 
  =
  \int_{T_0}^T \mathrm{d}T' \frac{\Delta(T';\bar\mu)}{T'}.
\end{equation}
In order to evaluate this we need to fix the pressure at a reference temperature $T_0$. We use the results in refs.\,\cite{Laine:2006cp,Laine:2015kra} at the lowest temperature in our temperature range, $T_0=140$\,GeV: $p(T_0)/T_0^4 = 11.173$.  The estimated uncertainty in this value is of order 1\%; we discuss this together 
with other systematic errors at the end of this section.

From $\Delta(T)$ and $p(T)$ we can obtain other thermodynamic functions:
the energy density $e/T^4 = \Delta + 3p/T^4$, the entropy density $s=p'= (e+p)/T$, the heat capacity $C_V/T^3 = e'/T^3 = 7\Delta + T\Delta' + 12p/T^4$, the speed of the sound squared $c_s^2 = p'/e'$ and the equation of the state parameter $w=p/e$.  The quantities involving $\Delta'$ can in principle be calculated directly from \eqr{delta} by numerical differentiation.  However, we obtain a numerically more robust result by realizing that to the parametric accuracy needed
\begin{equation}
  \frac{\mathrm{d}}{\mathrm d T} \frac{\langle\phi^\dagger\phi\rangle}{T}
  \approx \frac{\mathrm{d} y}{\mathrm d T} \frac{\mathrm d}{\mathrm d y} \frac{\langle\phi^\dagger\phi\rangle}{T}
  = -\frac{\mathrm d y}{\mathrm d T} \susk.
\end{equation}
The direct numerical differentiation of $\langle\phi^\dagger\phi\rangle$ gives comparable results, but because it was obtained through continuum extraplation of spline interpolations the derivative of it becomes somewhat jagged.

The final results are collected in figure \ref{fig:thermo},
with error bands which combine quadratically the statistical errors and the variation in the renormalization scale $\bar\mu = (0.5T \ldots 2)\pi T$.  The renormalization scale dominates the error bands except in quantities where $\Delta'$ contributes, where the statistical errors are of comparable magnitude.  
The results show similar features than the earlier ones published in ref.\,\cite{Laine:2015kra}, but with improved accuracy and reliability, because the Higgs condensate and susceptibility have been properly extrapolated to the continuum.
The tabulated numerical results can be downloaded from \cite{sm_data}.

Figure \ref{fig:thermo} clearly shows that the cross-over region, defined as the temperature range where the thermodynamic variables deviate from their (perturbative) hot or cold phase behaviour, is narrow, between $\sim 157$ and 162\,GeV.  This is precisely the region where perturbation theory does not converge and lattice simulations are needed.  The variation of the thermodynamic quantities across the cross-over is very mild, only at few per cent level.  This is a natural consequence of the fact that only a couple of degrees of freedom of the Standard Model are strongly sensitive to the cross-over. The behaviour is markedly different from the QCD cross-over, see e.g.\,\cite{Bazavov:2015rfa} for a recent review.

In addition to the statistical errors and renormalization scale ambiguity shown in figure \ref{fig:thermo}, there are theoretical uncertainties that can be estimated to be at percent level \cite{Laine:2015kra}, especially affecting the energy density and the pressure.  As an indication of these a 1\% bar is shown in figure \ref{fig:thermo}, and it clearly dominates over the error bands of $e$ and $p$.
This uncertainty mostly pertains to absolute normalization in these quantities, whereas the shape of the curves can be expected to be preserved to much better accuracy.  In particular, the characteristic features of the thermodynamic functions around the cross-over are mostly due to the evolution of $\langle\phi^\dagger\phi\rangle$ and hence more accurately determined.

\section{Screening masses}
\label{sec:screening}

The screening masses of the Higgs and SU(2) gauge fields are direct probes of the ``softening'' of the physics at the cross-over.   We measure the Higgs, $W^3$ and hypercharge $B$ spatial correlation functions and extract the screening masses (inverse correlation lengths) as functions of the temperature.
The Higgs and $W^3$ correlation functions are measured with lattice operators corresponding to $H=\tr\Phi^\dagger\Phi$ and $W^3_i = \tr\Phi^\dagger iD_i \Phi\sigma_3$, projected to zero momentum.  For example, the Higgs correlation function is (here in units of lattice spacing $a$):
\begin{equation}
  C_{H}(z) = \frac{1}{N^3}\sum_{x_i,y_i,z'} \langle H(x_1,y_1,z') H(x_2,y_2, z'+ z)\rangle,
\end{equation}
where $x_i,y_i,z,z'$ are $(x,y,z)$ coordinates compatible with the periodic boundary conditions.
The local operators corresponding to $W^+$ and $W^-$ (linear combinations of $\tr\Phi^\dagger iD_i\Phi\sigma_a$ with $a=1,2$) are not gauge invariant under U(1) and we do not measure them here.  To enhance the signal, we use lattice operators which are smeared and blocked recursively 3 times, as described in ref.\,\cite{su2xu1}.
\begin{figure}
  \includegraphics[scale=0.6]{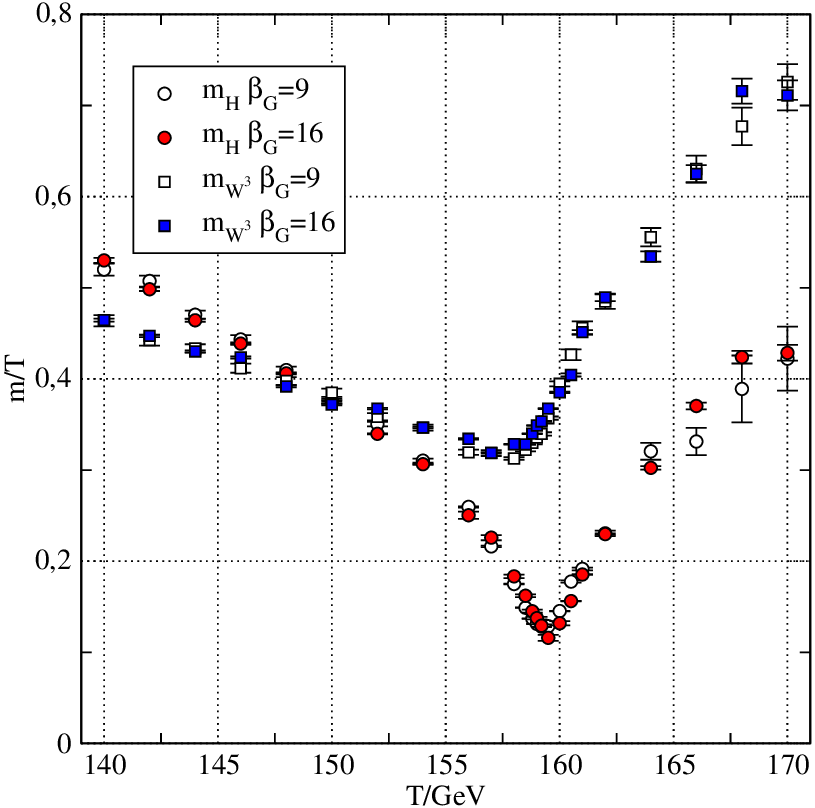}
  \caption{Higgs and $W^3$ screening masses, measured from
    $60^3$, $\beta_G=9$ and $120^3$, $\beta_G=16$ lattices.}
  \label{fig:mHmW}
\end{figure}
The screening masses extracted from the correlation functions are shown in figure \ref{fig:mHmW}.  We use the largest lattices at $\beta_G=9$ and $\beta_G=16$, with results which are consistent within the statistical errors.

In the broken phase, where $\phi^\dagger\phi$ has a large expectation value, the operators couple appropriately to the physical Higgs and $Z$-boson excitations.  At small temperatures the Higgs screening mass is larger than that of $W^3$, whereas near the cross-over temperature the Higgs becomes sensitive to the near-critical behaviour and the mass becomes small but still remains non-zero.  Even at its largest, the Higgs correlation length is smaller than $10/T$, which is substantially smaller than the largest lattice sizes $\sim 70$-$80/T$.

In the symmetric phase, the non-abelian gauge fields are confining, and the operators couple to bound states of two scalars.  The correlation functions become noisy and the screening masses increase rapidly.


The U(1) gauge field correlation function can be used to measure the $\gamma$-$Z$ mixing, i.e. the effective Weinberg angle.   We define the operator
\begin{equation}
  O_{\bf p}(z) = \sum_{x_1,x_2} \alpha_{12}(x_1,x_2,z)
  e^{i {\bf p}\cdot{\bf x}},
\end{equation}
where the sum is taken over the plane $(x_1,x_2)$, $\alpha_{ij}$ is the (non-compact) hypercharge U(1) plaquette \eqr{u1plaq} and ${\bf p}$ is a transverse momentum vector compatible with periodic boundary conditions:  $(p_1,p_2,p_3) = 2\pi/N (n_1,n_2,0)$ with integer $n_1$ and $n_2$. In our measurements we use the smallest non-vanishing momentum, with $|{\bf p}| = 2\pi/N$. At ${\bf p}=0$ the operator $O_{\bf p}$ vanishes, due to the periodic boundary conditions.  The correlation function
\begin{equation}
  G(z) = \frac{1}{N^3} \sum_t \langle O_{\bf p}(t) O^*_{\bf p}(z+t)\rangle
\end{equation}
has the long distance behaviour \cite{su2xu1}
\begin{equation}
  G(z) \rightarrow \frac{A_\gamma z}{2\beta_G} \frac{ap^2}{\sqrt{p^2+m_\gamma^2}}
  e^{-z \sqrt{p^2+m_\gamma^2}}
\end{equation}
where $m_\gamma$ is the photon screening mass and $A_\gamma$ gives the projection of the operator to the hypercharge U(1) field, in effect yielding the temperature-dependent effective mixing angle.  At tree level, $A_\gamma = 1$ in the symmetric phase and $A_\gamma = \cos^2\theta_W$ in the broken phase. 

\begin{figure}
  \includegraphics[scale=0.6]{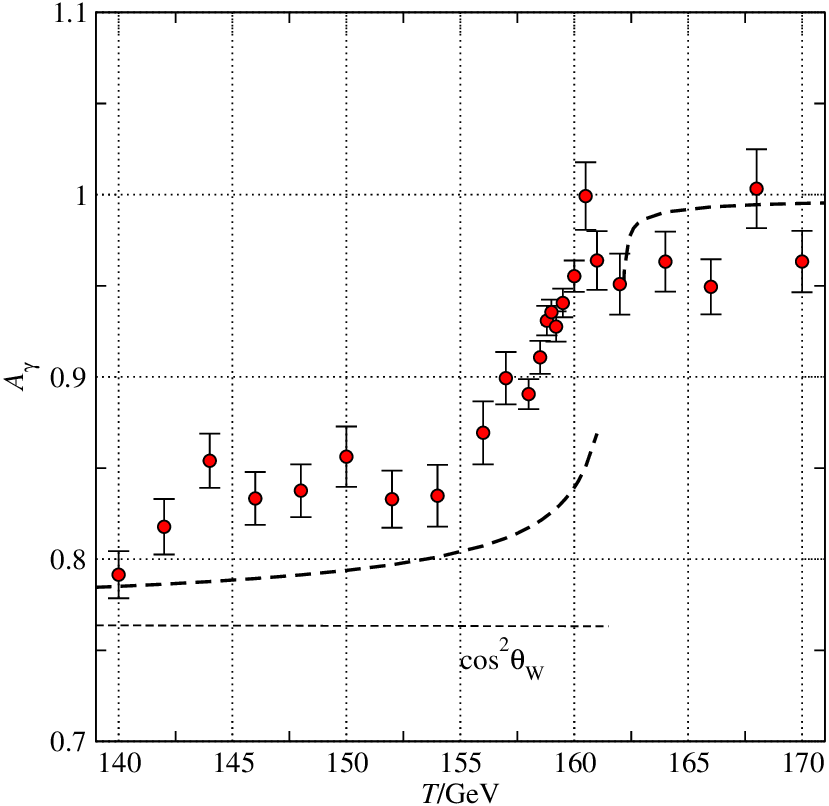}
  \caption{The effective $\gamma-Z$ mixing as a function of the temperature.  The dashed lines show the 1-loop perturbative results.}
  \label{fig:Ag}
\end{figure}

The photon screening mass $m_\gamma$ vanishes within our measurement accuracy at all temperatures.  The projection $A_\gamma$ is shown in figure \ref{fig:Ag} for $\beta_G=9$, $60^3$ lattice.  The measurement is noisy, but we can observe that $A_\gamma \approx 1$ in the symmetric phase down to the cross-over temperature, and it starts to decrase as the Higgs field expectation value grows at lower temperatures, slowly approaching the tree-level value. 

Beyond tree-level perturbative estimates for the behaviour of $A_\gamma$ can be obtained by calculating at 1-loop order the residue  of the $1/k^2$ pole in the $\langle B_i B_j\rangle$ correlator.  In the symmetric and broken phases one obtains \cite{su2xu1}
\begin{align}
  A_\gamma^{\rm symm.} &= 1 - \frac{z}{48\pi\sqrt{y}} \\
  A_\gamma^{\rm broken} &= \cos^2\theta_W \left(1+\frac{11}{12}\frac{g_3^2\sin^2\theta_W}{\pi m_W}\right)
\end{align}
where $m_W$ is the perturbative $W$ mass.  These expressions clearly anticipate the behaviour we observe on the lattice, although they diverge as $y\rightarrow 0\pm$.


\section{Conclusions}
\label{sec:conclusions}

We have accurately determined the Higgs field expectation value and its susceptibility across the Standard Model cross-over using lattice simulations of an effective 3-dimensional theory.  Defining the cross-over temperature by the maximum of the susceptibility, we obtain $T_c=159.6 \pm 0.1 \pm 1.5$\,GeV, where the first error is due to the statistical accuracy of the lattice computation and the second one is the estimated uncertainty of the effective theory approach \cite{generic,Laine:2015kra}.  Following the approach of Laine and Meyer~\cite{Laine:2015kra}, these results were 
used to obtain the behaviour of basic thermodynamic quantities, including energy density, pressure, heat capacity and the speed of sound, across the cross-over.  There is a well-defined cross-over region where thermodynamic quantities deviate from the low- or high-temperature behaviour.  This region is quite narrow, between 157 and 162\,GeV.  
The results are consistent with the standard picture of the electroweak cross-over: Higgs and $W$ modes become softer but not critical, and the U(1) field remains massless at all temperatures.

Overall our results are compatible with the analysis in ref.\,\cite{Laine:2015kra} using lattice data from ref.\,\cite{D'Onofrio:2014kta}.  Howeever, our results are significantly improved numerically:
we have much larger volumes with higher statistical accuracy, the data is extrapolated to the continuum and we include the U(1) field in the effective theory.  Thus, our results form an important consistency and reliability check of the earlier results.

For phenomenological applications the thermodynamic quantities here can be combined with existing low- \cite{Laine:2006cp} and high-temperature \cite{Gynther:2005dj} perturbative results.  This has been done in ref.\,\cite{Laine:2015kra} and we do not repeat this analysis here.

The errors are dominated by systematic uncertainties at per cent level, but since these are expected to be roughly constant at the temperature range considered, they would mostly shift the scale of the results without changing their behaviour.
The cross-over is clearly visible in distinct features in the heat capacity, the speed of sound and the interaction measure.  The relative magnitude of these features is small, only at few \% level, which already follows from the fact that only few degrees of freedom of the Standard Model are sensitive to the early stages of the transition.  

In ref.\,\cite{D'Onofrio:2014kta} the central value of the cross-over temperature was reported to be $159.5$\,GeV, in practice equal to our continuum result here.  This work was done without the U(1) field and only at one lattice spacing, $\beta_G=9$, using relatively small statistics.  However, 
one cannot make the conclusion that the effects of the U(1) field can be neglected in the transition: from figure \ref{fig:phi22vol} we see that at $\beta_G=9$ the maximum of the susceptibility is at $\approx 159.0$\,GeV, which is a numerically significant deviation from the continuum result.  Thus, the apparent excellent agreement is partly coincidental, due to a different definition of the cross-over temperature and limited accuracy in \cite{D'Onofrio:2014kta}.  Nevertheless, we can estimate that the presence of the U(1) field in the effective theory shifts the cross-over temperature $\lsim 0.5$\,GeV.


\begin{acknowledgments}
  We thank Mikko Laine for discussions.  This work has been supported by the Magnus Ehrnrooth Foundation (MD) and by the Finnish  Academy through grants 1134018 and 1267286. Part of the numerical work has been performed using the resources at the Finnish IT Center for Science, CSC.
\end{acknowledgments}

\appendix
\section{Interpolating function for the susceptibility}
\label{app:fit}

In this subsection we describe the method used for the interpolation of the susceptibility measurement at different lattice volumes and $\beta_G$-values in section \ref{sec:condensate}.  Instead of the temperature, it is convenient to do the interpolation in terms of the 3-dimensional variable $y=m_3^2/g_3^2$, which is actually the variable used internally throughout our analysis.

The functional form of the fit is motivated by the sharp susceptibility peak near $y=0$ ($T=162.1$\,GeV), approaching almost flat behaviour away from the peak.  Thus, we construct a differentiable function which approaches power series in $1/y$ on both sides of the cross-over.  For our data we obtain very good results with the function 
\begin{equation}
  \chi_{\rm fit}(y,\beta_G) = \frac{(\sum_{i=0}^{n_1} c_i s^{i}) E_1(y) +
    (\sum_{i=0}^{n_2} d_i s^{i}) E_2(y)}{E_1(y) + E_2(y)},
\end{equation}
where 
\begin{align*}
  y_s &= y-a_1\\
  s &= 1/\sqrt{(y_s^2 + a_2} \\
  E_1(y) &= 1/E_2(y) = e^{-a_3 y_s}
\end{align*}
Here $a_1 \ldots a_3$, $c_0 \ldots c_{n_1}$ and $d_0 \ldots d_{n_2}$ are fit parameters, $s$ corresponds to a regulated $|y|^{-1}$ and the exponential factors $E_{1,2}$ are introduced to smoothly switch between the $c$-series on the low temperature side and the $d$-series on the high temperature side of the cross-over.  Parameter $a_1$ corresponds roughly to the location of the peak, and $a_2>0$ regulates the singular behaviour at $y_s\rightarrow 0$.

Using $n_1=3$, $n_2=2$ we obtain a good fit to all measurements ($\chi^2/{\rm d.o.f.} = 13/16$, $20/16$ and $10/16$ for the largest volumes at $\beta_G=6$, $9$ and $16$, respectively), as shown in figure \ref{fig:phi22}.  Adding higher order terms does not improve the fit, and the final function remains practically identical, with variation remaining well within the statistical errors.

The fit function is not unique, and the fit parameters have no direct physical meaning.  The function only serves to interpolate the measurements and cannot be extended beyond the range of data shown here.  Naturally, significantly more accurate data or extended temperature range would require a more general fit function or some other means of interpolation.  The error propagation is done with jackknife analysis.

\bibliography{u1xsu2}

\end{document}